\documentclass[9pt,conference]{IEEEtran}
\usepackage{amssymb,amsthm,amsmath,array}
\usepackage{graphicx}
\usepackage[caption=false,font=footnotesize]{subfig}
\usepackage{xspace}
\usepackage[sort&compress, numbers]{natbib}
\usepackage{stmaryrd}
\usepackage{xcolor}
\usepackage{mathtools}
\usepackage{float}
\usepackage{textcomp}
\usepackage{color}
\usepackage{booktabs}
\usepackage{enumitem}
\usepackage{multirow}
\usepackage[normalem]{ulem} % Provides \sout for strike-through

\usepackage[nolist]{acronym}
\usepackage{hyperref}
\usepackage{cleveref}
\newcommand{\data}{f}
\newcommand{\reconst}{u}
\newcommand{\fourier}{\mathcal{F}}
\newcommand{\pupilOp}{\mathcal{M}}

\newcommand{\na}{\text{NA}}
\newcommand{\wavelength}{\lambda}

\begin{acronym}
  \acro{fpm}[FPM]{Fourier Ptychographic Miscroscopy}
  \acro{cnn}[CNN]{Convolutional Neural Networks}
  \acro{na}[NA]{Numerical Aperature}
  \acro{la}[LA]{Limiting Aperature}
\end{acronym}

\begin{document}
\title{Direct Image Classification from Fourier Ptychographic Microscopy Measurements without  Reconstruction}
\author{\IEEEauthorblockN{
        Navya Sonal Agarwal\IEEEauthorrefmark{1},
        Jan Philipp Schneider\IEEEauthorrefmark{2}, 
        Kanchana Vaishnavi Gandikota\IEEEauthorrefmark{2}, \\
        Syed Muhammad Kazim\IEEEauthorrefmark{3}, 
        John Meshreki\IEEEauthorrefmark{3}, 
        Ivo Ihrke\IEEEauthorrefmark{3},
        and Michael Moeller\IEEEauthorrefmark{2}
    }
    \IEEEauthorblockA{
     \IEEEauthorrefmark{1} Kalinga Institute of Industrial Technology, India. \\
     \IEEEauthorrefmark{2} Chair for Computer Vision, University of Siegen, Germany.\\ 
     \IEEEauthorrefmark{3} Chair for Computational Sensing and Communications Engineering, University of Siegen, Germany.\\
        }
}

\maketitle
\begin{abstract}
  The computational imaging technique of Fourier Ptychographic Microscopy (FPM) enables high-resolution imaging with a wide field of view and can serve as an extremely valuable tool, e.g. in the classification of cells in medical applications. However, reconstructing a high-resolution image from tens or even hundreds of measurements is computationally expensive, particularly for a wide field of view. Therefore, in this paper, we investigate the idea of classifying the image content in the FPM measurements directly without performing a reconstruction step first. We show that \ac{cnn} can extract meaningful information from measurement sequences, significantly outperforming the classification on a single band-limited image (up to 12 \%) while being significantly more efficient than a reconstruction of a high-resolution image. Furthermore, we demonstrate that a learned multiplexing of several raw measurements allows maintaining the classification accuracy while reducing the amount of data (and consequently also the acquisition time) significantly. 
  \end{abstract}

\section{Introduction}
\ac{fpm} \cite{zheng_2013_widefield} has become increasingly popular in recent years due to its ability to reconstruct super-resolved amplitude and phase images with a wide field of view. This opens up numerous image enhancement and application possibilities in areas like biology and medicine. 
\ac{fpm}  can be implemented with low hardware costs by replacing the illumination unit of the microscope with an LED array to illuminate the specimen from different lighting angles. However, considering that hundreds of images of 10 or more megapixels need to be combined to a high-resolution gigapixel image via an optimization method, \ac{fpm}  requires large computational resources.  

In this paper, we propose to mitigate computational demands in applications where the image content needs to be categorized by directly feeding the \textrm{measurements}, i.e. the images recorded using different illumination directions, into a \ac{cnn}. We demonstrate that the resulting classification significantly improves over the accuracy obtained on a single on-axis brightfield illumination image only. Our approach can be particularly attractive in applications where small regions of interest are scattered in the field of view, e.g. in cytopathology: It allows identifying regions of interest which can then be reconstructed and further examined by smaller measurement patches. While reconstructions on small measurement patches are common in \ac{fpm} (c.f. \cite{zheng_2013_widefield}), our work is, to the best of our knowledge, the first to consider the direct classification of direct measurement data. 

Furthermore, we propose to incorporate \textit{multiplexing}~\cite{tian_2014, kellman:2019}, a technique that combines measurements by illuminating the probe with multiple LEDs simultaneously. We demonstrate that although one has to expect a loss of image quality in \ac{fpm} reconstruction, the direct classification from raw measurements remains remarkably unaffected by the multiplexing, allowing to reduce the number of recorded images by up to $60\%$ without significant loss in classification accuracy.

\section{Background}
\begin{figure}[t]
    \centering
    \includegraphics[width=0.51\linewidth]{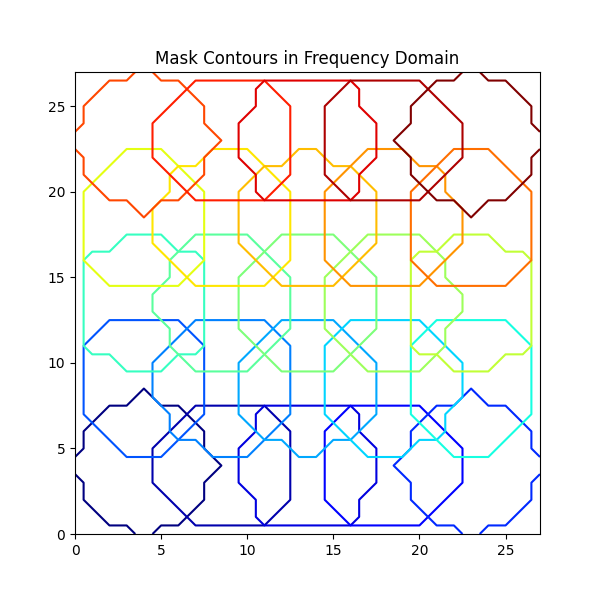}
    \includegraphics[width=0.45\linewidth]{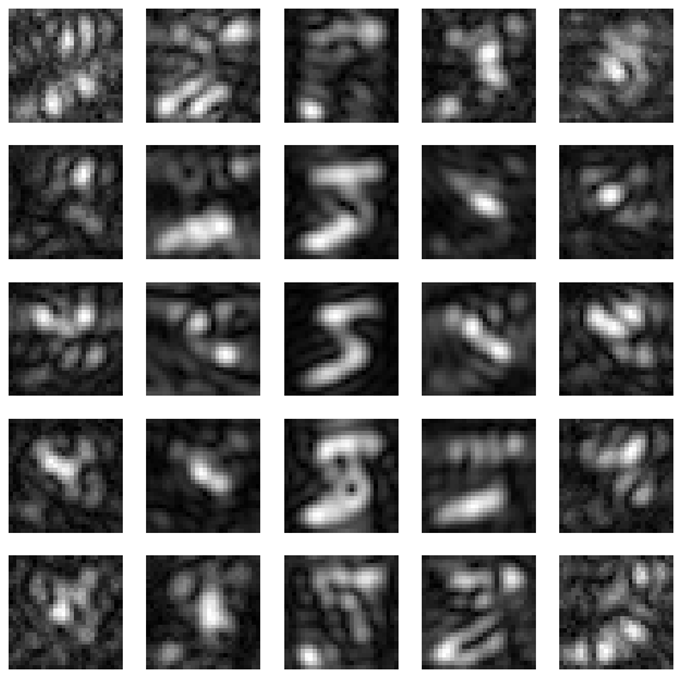}
    \caption{Illustration of FPM measurements: Each colored contour represents one cropped spectrum in the frequency domain (left) and the corresponding band-limited FPM measurement (right).}
    \label{fig:data}
    \vspace{-0.5cm}
\end{figure}

As the core idea of \ac{fpm} is to record individual components of the specimen's diffraction spectrum by changing the illumination angle with several LEDs, 
the data formation model \cite{zheng_2013_widefield} 
\begin{align}
    \data^k = |\fourier^{-1}\pupilOp^k\fourier\reconst|+ \text{noise},
    \label{eq:fpm}
\end{align}
can be used to represent the image formation of a single band-limited measurement $\data^{k} \in \mathbb{R}^{W \times H}$. Further, $\reconst \in \mathbb{C}$ denotes the true high-resolution specimen to reconstruct, $\fourier$ denotes the Fourier transform, $\pupilOp^k$ is an operator that models the pupil function and the \ac{la}. 
In the simplest case, the operator $\pupilOp^k$ can be modeled as a cropping operation that sets all values outside of a circle with LED-position dependent center and a radius proportional to $(2 \pi/ \lambda) \na $ to zero, where $\wavelength$ denotes the illumination wavelength, and $\na$ the \ac{na} of the objective lens.
In the absence of a physical setup, the radius and position of the cropping operator $\pupilOp^k$ can be chosen arbitrarily, allowing us to simulate various \ac{na} sizes, led positions, and alignments. An example of simulated FPM measurements is provided in \cref{fig:data}, which illustrates the cropping operation at different regions in Fourier space corresponding to LED positions, and the respective raw FPM measurements.

The data formation model \eqref{eq:fpm} gives rise to an inverse problem frequently being solved by optimizing
\begin{align}
\label{eq:optimization}
  \arg\min_\reconst \ \ \left( \sum_{k=1}^K  H(\data^k, |\fourier^{-1}\pupilOp^k\fourier\reconst|) + \alpha R(\reconst) \right),
\end{align}
for a suitable measure of discrepancy $H$ (e.g. an $\ell^1$ norm) and a suitable regularizer $R$, e.g. the total variation of $\reconst$. Considering that each measurement $\data^k$ commonly is an image of several megapixels and that $K$ commonly is in the order of 100, Eq.~\eqref{eq:optimization} is expensive to solve.

To at least reduce the measurement time, the original \ac{fpm} formulation of \cite{zheng_2013_widefield} was extended in \cite{tian_2014} by using multiplexing, as well as learning the specific LED pattern and illumination intensities of the multiplexing in \cite{kellman2019physics}. Such approaches can be modeled by a weighted linear combination of the measurements that one would have obtained from a single LED illumination, i.e.
\begin{align}
    \data^k = \sum_l \beta^k_l |\fourier^{-1} \pupilOp^k_l\fourier\reconst|+ \text{noise},
    \label{eq:fpm_multiplexing}
\end{align}
for non-negative weights $\beta^k_l$ representing the illumination intensity and $\pupilOp^k_l$ representing the operators arising from different (simultaneously active) LEDs.

\section{Numerical Experiments}
Considering the high computational demands of first solving \eqref{eq:optimization} and subsequently feeding the reconstructed high-resolution image into a classifier, we investigate the idea of training a classifier on direct measurement data, i.e., on the collection of all $\data^k$ stacked into an image with $K$ channels.

We perform synthetic experiments using three popular image recognition datasets, each of which have 10 classes: MNIST \cite{lecun1998gradient} with grey scale images of digits of resolution $28\times28$ pixels, and the color image datasets CIFAR 10 \cite{krizhevsky2009learning}  and Imagenette \cite{imagenette} with images resized to resolution $100\times100$ pixels.  We simulate FPM measurements of images from these datasets by applying \eqref{eq:fpm} considering $25$ crops in the Fourier space for different values of \ac{na}. We conduct experiments using Resnet18 \cite{he2016deep} comparing the following scenarios: (i) providing only the central measurement, denoted by CC (central crop) in \cref{tab:results1}, (ii) providing a stack of all the low-resolution measurements, denoted by SC (stacked crop), (iii) providing the reconstructed images obtained through a gradient descent-based energy minimization scheme using TV regularization, denoted by R, and (iv) providing the ground truth images as an upper bound (UB).

We modify the first layer of the network to reflect the increased number of measurements for both the stacked crop and the multiplexed setting.  The networks are trained with a cross-entropy loss using the Adam optimizer \cite{kingma2014adam} with learning rates of $1e-3$ for MNIST, $1e-4$ for CIFAR 10 and Imagenette for 20 epochs. We train networks for CIFAR10 and MNIST from scratch and finetune a pretrained Resnet on Imagenette. We report the classification accuracy of the models at the final epoch.

The results of this experiment are provided in \cref{tab:results1}. While using a single FPM measurement (CC) alone results in comparatively low accuracy, using all the 25 low-resolution measurements improves the accuracy by 1-1.5\% in MNIST, 10-12\% in CIFAR10 and 0.5-6\% in Imagenette. In our simple experiment, the accuracy of a proper reconstruction (R) almost reaches the upper bound as we are using the same operator for simulation and reconstruction, and assume the prior knowledge of the image to be reconstructed having a zero phase. The fact that the stacked raw-data classification remains inferior to a full reconstruction is to be expected, as learning the backward model of FPM in a purely data-driven way using similar zero-shot supervised formulations is empirically observed to be a challenging task \cite{zhao2023transformer}. However, as the reconstruction involves significant computational costs, pre-classification using the direct measurements, as proposed in our SC setting, is very attractive to determine regions of interest.  
Variations in the cropping radius of the pupil (corresponding to varying limiting apertures) yield surprisingly inconclusive results with SC-accuracies not always being monotonically increasing with increasing pupil radius.

Additionally, we considered simulating multiplexing through multiple-LED illumination by learning a linear combination of measurements implemented using a $1\times1$ convolution layer whose non-negative weights are learned along with the classification network weights.  The results of this experiment are shown in \cref{tab:linearcomb}, where we simulated multiplexing up to 25 different LEDs (with different intensities) into 5, 10, 15 and 20 measurements, which are provided as inputs to the classification network. The results show only a small reduction in accuracy for low-resolution datasets ($\sim3\%$ on CIFAR10) even when the number of multiplexed measurements is reduced to 5 in comparison to using all 25 band-limited measurements. With higher number of multiplexed measurements ($\geq10$), the classifier accuracy is maintained. For the Imagenette dataset, we trained the network for 40 epochs instead of 20, as the learned weights for multiplexing were not trained well enough at 20 epochs. For a fair comparison, we repeated the SC and CC also for 40 epochs, which, however, resulted in lower accuracies due to overfitting. Even on Imagenette, we observe that the learned multiplexing results in improved accuracies compared to single measurements, with only a small drop in accuracy,  indicating the benefit of learning the illumination pattern to record fewer measurements and further reduce the computational demands.

\begin{table}[h!]
\centering
\resizebox{0.9\linewidth}{!}{
\begin{tabular}{|c|c|c|c|c|c|}
\hline
\multirow{2}{*}{\textbf{Dataset}} &  {\textbf{Pupil}} & \multicolumn{4}{c|}{\textbf{Accuracy (\%)}} \\ 
\cline{3-6}              &   \textbf{Radius}                                 & \textbf{CC} & \textbf{SC} & \textbf{R} & \textbf{UB} \\ \hline
\multirow{3}{*}{\vtop{\hbox{\strut \textbf{MNIST}}\hbox{\strut $28\times28$}}}                                       
       & 3                                  & 97.76       & 99.33                               & \multirow{3}{*}{99.42} & \multirow{3}{*}{99.47}                        \\ \cline{2-4}
                                   &                                     4                                  & 98.59       & 99.35       &                         &                         \\ \cline{2-4}
                                   &                                    5                                  & 99.06       & 99.02       &                         &                         \\ \hline
\multirow{3}{*}{\vtop{\hbox{\strut \textbf{CIFAR10}}\hbox{\strut~$32\times32$}}}                   
                                   & 3                                  & 60.70       & 72.32       & \multirow{3}{*}{87.27} & \multirow{3}{*}{95.52}         \\ \cline{2-4}                        
                                   & 4                                  & 63.38      & 75.83       &                         &                         \\ \cline{2-4}
                                   & 5                                  & 65.64       & 75.14       &                         &                         \\ \hline
\multirow{3}{*}{\vtop{\hbox{\strut \textbf{Imagenette}}\hbox{\strut $100\times100$}}}                      & 15                                 & 75.08      & 81.2      & \multirow{3}{*}{91.20} & \multirow{3}{*}{93.52} \\ \cline{2-4}                            & 20                                 & 77.73      & 76.99       &                         &                         \\ \cline{2-4}   & 25                                 &  75.62    & 76.10       &                         &                         \\ \hline
\end{tabular}}
    \vspace{0.05cm}
\caption{Comparison of accuracies of a Resnet18 trained on the central measurement (CC), a stacking of all measurements (SC), on reconstructed images (R) and ground truth (UB).% for a single run with a fixed seed 0.
\label{tab:results1}}
\vspace{-0.8cm}
\end{table}
\begin{table}[h!]
    \centering
    \resizebox{\linewidth}{!}{
    \begin{tabular}{cccccccc}
    \toprule
\multirow{2}{*}{\textbf{Dataset}} &  \multirow{2}{*}{\textbf{Radius}}  & \multirow{2}{*}{\textbf{SC}} &  \multirow{2}{*}{\textbf{CC}}&\multicolumn{4}{c}{\textbf{Number of multiplexed measuremets}}\\\cmidrule{5-8}
&&&&\textbf{5}&\textbf{10} & \textbf{15} & \textbf{20} \\
\midrule
MNIST & 5 &99.25 & 99.06&99.22 &99.37 & 99.50 & 99.28\\
CIFAR10 & 5  &75.14&65.64&72.32&75.37&74.88&75.53\\
Imagenette & 15&81.2&75.08 &76.99&79.21&78.42&81.07\\
         \bottomrule
    \end{tabular}}
    \vspace{0.05cm}
    \caption{Classification accuracies using all FPM measurements vs. multiplexed measurements}
%\VG{number in purple is the result with 20 epochs}}%\JPS{Why is CIFAR10-5 different than in the above table?} \VG{older value without fixed seed, replaced now}}
    \label{tab:linearcomb}
\end{table}
\vspace{-0.7cm}
\section{Conclusions}
In this paper, we investigated the utility of direct FPM measurements in deep learning-based classification. When the number of band-limited measurements is increased, a performance increase of up to 12 \% can be observed compared to a center crop (SC), highlighting the ability of the networks to extract information from the additional frequency spectrum covered. We also found that the number of FPM measurements can be compressed and reduced using a learned linear combination with a negligible loss in classification accuracy compared to the full measurement. These results indicate the potential of directly using FPM measurements in extracting potential regions of interest in Gigapixel scale images, reducing the computational overhead associated with their analysis.

\section{Acknowledgments}
This work was supported by the German Research Foundation (DFG) under grants FOR 5336 (IH 114/2-1 and MO 2962/11-1).

\bibliographystyle{IEEEtran}

\end{document}